\documentclass{gdfp}

\usepackage{natbib}
\usepackage{rotating}

\usepackage{graphicx}
\usepackage{amsmath,amsthm,bm,mathrsfs}

\renewenvironment{proof}[1][Proof]{\noindent\textit{#1. } }{\hfill$\square$}

 \newtheoremstyle{theorem}{6pt}{6pt}{\rm}{}{\sffamily}{ }{ }{}
 \theoremstyle{theorem}
\newtheorem{theorem}{\sc Theorem}[section]

 \newtheoremstyle{algorithm}{6pt}{6pt}{\rm}{}{\sffamily}{ }{ }{}
 \theoremstyle{algorithm}

 \newtheoremstyle{lemma}{6pt}{6pt}{\rm}{}{\sffamily}{ }{ }{}
 \theoremstyle{lemma}
 \newtheorem{lemma}{\sc Lemma}[section]

\newtheoremstyle{case}{6pt}{6pt}{\rm}{}{\sffamily}{. }{ }{}
 \theoremstyle{case}

 \newtheoremstyle{statement}{6pt}{6pt}{\rm}{}{\sffamily}{ }{ }{}
\theoremstyle{statement}

 \newtheoremstyle{corollary}{6pt}{6pt}{\rm}{}{\sffamily}{ }{ }{}
 \theoremstyle{corollary}

  \newtheoremstyle{definition}{6pt}{6pt}{\rm}{}{\sffamily}{ }{ }{}
 \theoremstyle{definition}

\newtheoremstyle{example}{6pt}{6pt}{\rm}{}{\sffamily}{ }{ }{}
\theoremstyle{example}

\newtheoremstyle{remark}{6pt}{6pt}{\rm}{}{\sffamily}{ }{ }{}
\theoremstyle{remark}
\newtheorem{remark}{\sc Remark}[section]

\newtheoremstyle{approximation}{6pt}{6pt}{\rm}{}{\sffamily}{ }{ }{}
\theoremstyle{approximation}

\newtheoremstyle{scheme}{6pt}{6pt}{\rm}{}{\sffamily}{ }{ }{}
\theoremstyle{scheme}

\newtheoremstyle{Algorithm}{6pt}{6pt}{\rm}{}{\sffamily}{ }{ }{}
\theoremstyle{Algorithm}

\newtheoremstyle{Assumption}{6pt}{6pt}{\rm}{}{\sffamily}{ }{ }{}
\theoremstyle{Assumption}

\newtheoremstyle{proposition}{6pt}{6pt}{\rm}{}{\sffamily}{ }{ }{}
\theoremstyle{proposition}

\newtheoremstyle{hypo}{6pt}{6pt}{\rm}{}{\sffamily}{ }{ }{}
 \theoremstyle{hypo}

  \newtheoremstyle{Step}{6pt}{6pt}{\rm}{}{}{ }{ }{}
 \theoremstyle{Step}

\numberwithin{equation}{section}

\newcommand{\N}{{\rm I\! N}}
\newcommand{\R}{{\rm I\! R}}

\newcommand{\QED}{\hfill $\Box$}

\newcommand{\Prob}{\mathbb{P}}

\begin{document}

\title{Multivariate high-frequency financial data via semi-Markov processes }
\author{ {\sc Guglielmo D'Amico}\\[2pt]
Dipartimento di Farmacia, Universit\`a "G. d'Annunzio" di Chieti-Pescara,\\
 via dei Vestini 31, 66013, Chieti, Italy.\\[6pt]
{\sc Filippo Petroni}\\[2pt]
Dipartimento di Scienze Economiche ed Aziendali, Universit\`a di Cagliari,\\
 09123 Cagliari, Italy.\\[6pt]
}
\pagestyle{headings}
\markboth{G. D'AMICO \& F. PETRONI}{\rm HIGH-FREQUENCY DATA VIA SEMI-MARKOV PROCESS}
\maketitle

\begin{abstract}
{In this paper we propose a bivariate generalization of a weighted indexed semi-Markov chains to study the high frequency price dynamics of traded stocks. We assume that financial returns are described by a weighted indexed semi-Markov chain model. We show, through Monte Carlo simulations, that the model is able to reproduce important stylized facts of financial time series like the persistence of volatility and at the same time it can reproduce the correlation between stocks. The model is applied to data from Italian stock market from 1 January 2007 until the end of December 2010.}{Financial market; semi-Markov chain; Bivariate processes}\end{abstract}


\section{Introduction}

Semi-Markov processes (SMP) are a wide class of stochastic processes which generalize at the same time both Markov chains and renewal processes. The main advantage of SMP is that they allow the use of whatever type of waiting time distribution for modeling the time to have a transition from one state to another one. On the contrary, Markovian models have constraints on the distribution of the waiting times in the states which should be necessarily represented by memory-less distributions (exponential or geometric for continuous and discrete time cases respectively). This major flexibility has a price to pay: the parameters to be estimated are more numerous.\\
In three recent papers D'Amico and Petroni (2011, 2012a, 2012b) we showed that returns of stocks from financial market are better represented by semi-Markov processes than by a simple Markov chain. In particular, we have showed that volatility clustering, one of the main stylized fact of financial market, is almost exactly reproduced by an indexed semi-Markov process. More important, in our models, the volatility autocorrelation is obtained endogenously without introducing external or latent auxiliary stochastic processes.
To improve further our previous results, in this work, we propose a bivariate model where the indexed semi-markov process is used to reproduce simultaneously two stocks. 

The database used for the analysis is made of high frequency tick-by-tick price data from 20 stocks in Italian market from the first of January 2007 until end of December 2010. From prices we then define returns at one minute frequency.

\section{The Weighted-Indexed Semi-Markov Model} 
In this section we describe the weighted-indexed semi-Markov model that is able to represent higher-order dependencies between successive observations of a state variable. One way to increase the memory of the process is by using high-order semi-Markov processes as defined in Limnios and Opri\c{s}an (2003) and more recently revisited and extended in a discrete time framework in D'Amico et al. (2012). A more parsimonious model, named indexed semi-Markov chain (ISMC) model, has been defined by D'Amico and Petroni (2011), and there, it is showed that it describes appropriately important empirical regularities of financial time series such as the first passage time distribution and the autocorrelation function. A further improvement of the ISMC model was proposed in D'Amico and Petroni (2012b) named Weighted-Indexed Semi-Markov Chain (WISMC) model which allows the possibility of reproducing long-term dependence in the stock returns in a very efficient way.\\
\indent Let us describe briefly the WISMC model. We assume that the value of the financial asset under study is described by the time varying asset price $S(t)$. The return at time $t$ calculated over a time interval of length $1$ is defined as $\frac{S(t+1)-S(t)}{S(t)}$. The return process changes value in time, then we denote by $\{J_{n}\}_{n\in \N}$ the stochastic process with finite state space $E=\{1,2,...,s\}$ and describing the value of the return process at the $n$-th change.\\
\indent Let us consider the stochastic process $\{T_{n}\}_{n\in \N}$ with values in $\N$. The random variable $T_{n}$ describes the time in which the $n$-th change of the return process occurs.\\
\indent Let us consider also the stochastic process $\{U_{n}^{\lambda}\}_{n\in \N}$ with values in $\R$. The random variable $U_{n}^{\lambda}$ describes the value of the index process at the $n$-th transition.\\
\indent In reference D'Amico (2011), the process $\{U_{n}\}$ was defined as a reward accumulation process linked to the Markov Renewal Process $\{J_{n},T_{n}\}$; in D'Amico and Petroni (2011) the process $\{U_{n}\}$ was defined as a moving average of the reward process. In D'Amico and Petroni (2012b) motivated by the application to financial returns, a more flexible index process was defined:
\begin{equation}
\label{funcrela}
U_{n}^{\lambda}=\sum_{k=0}^{n-1}\sum_{a=T_{n-1-k}}^{T_{n-k}-1}f(J_{n-1-k},a,\lambda),
\end{equation}
where $f:E\times \N \times \R \rightarrow \R$ is a Borel measurable bounded function and $U_{0}^{\lambda}$ is known and non-random.\\
\indent The process $U_{n}^{\lambda}$ can be interpreted as an accumulated reward process with the function $f$ as a measure of the weighted rate of reward per unit time. The function $f$ depends on the current time $a$, on the state $J_{n-1-k}$ visited at current time and on the parameter $\lambda$ that represents the weight.\\
The application of the model requires the choice of a specific functional form of $f$.\\
\indent The WISMC model is completely specified once a dependence structure between the variables is considered. Toward this end, the following assumption was done:
\begin{equation}
\label{kernel}
\begin{aligned}
& {\Prob}[J_{n+1}=j,\: T_{n+1}-T_{n}\leq t |\sigma(J_{h},T_{h},U_{h}^{\lambda}),\, h=0,...,n, J_{n}=i, U_{n}^{\lambda}=v]\\
& =\mathbb{P}[J_{n+1}=j,\: T_{n+1}-T_{n}\leq t |J_{n}=i, U_{n}^{\lambda}=v]:=Q_{ij}^{\lambda}(v;t),
\end{aligned}
\end{equation}
\noindent where $\sigma(J_{h},T_{h},U_{h}^{\lambda}),\, h\leq n$ is the natural filtration of the three-variate process.\\
\indent The matrix of functions ${\bf Q}^{\lambda}(v;t)=(Q_{ij}^{\lambda}(v;t))_{i,j\in E}$ is called $\emph{weighted-indexed}$ $\emph{semi-Markov}$ $\emph{kernel}$.\\
\indent The joint process $(J_{n},T_{n})$ depends on the process $U_{n}^{\lambda}$, the latter acts as a stochastic index. Moreover, the index process $U_{n}^{\lambda}$ depends on $(J_{n},T_{n})$ through the functional relationship $(\ref{funcrela})$.\\
\indent Observe that if 
\[
\mathbb{P}[J_{n+1}=j,\: T_{n+1}-T_{n}\leq t |J_{n}=i, U_{n}^{\lambda}=v]=\mathbb{P}[J_{n+1}=j,\: T_{n+1}-T_{n}\leq t |J_{n}=i]
\]
\noindent for all values $v\in \R$ of the index process, then the WISMC kernel degenerates in an ordinary semi-Markov kernel and the WISMC model becomes equivalent to classical semi-Markov chain model.\\
\indent The triple of processes $\{J_{n}, T_{n}, U_{n}^{\lambda}\}$ describes the behavior of the system only in correspondence of the transition times $T_{n}$. To describe the behavior of our model at whatever time $t$ which can be a transition time or a waiting time, we need to define additional stochastic processes.\\
\indent Given the three-dimensional process $\{J_{n}, T_{n}, U_{n}^{\lambda}\}$ and the weighted indexed semi-Markov kernel ${\bf Q}^{\lambda}(v;t)$, we define by
\begin{equation}
\label{stocproc}
\begin{aligned}
& N(t)=\sup\{n\in \mathbb{N}: T_{n}\leq t\};\\
& Z(t)=J_{N(t)};\\
& U^{\lambda}(t)=\sum_{k=0}^{N(t)-1+\theta}\,\,\sum_{a=T_{N(t)+\theta -1-k}}^{(t\wedge T_{N(t)+\theta-k})-1}f(J_{N(t)+\theta-1-k},a,\lambda),
\end{aligned}
\end{equation}
where $\theta =1_{\{t>T_{N(t)}\}}$.\\
\indent The stochastic processes defined in $(\ref{stocproc})$ represent the number of transitions up to time $t$, the state of the system (price return) at time $t$ and the value of the index process (weighted moving average of function of price return) up to $t$, respectively. We refer to $Z(t)$ as a weighted indexed semi-Markov process.\\
\indent The process $U^{\lambda}(t)$ is a generalization of the process $U_{n}^{\lambda}$ where time $t$ can be a transition or a waiting time. It is simple to realize that if $t=T_{n}$ we have that $U^{\lambda}(t)=U_{n}^{\lambda}$.\\  
\indent Let 
$$
p_{ij}^{\lambda}(v):= \mathbb{P}[J_{n+1}=j|J_{n}=i,U_{n}^{\lambda}=v],
$$ 
be the transition probabilities of the embedded indexed Markov chain. It denotes the probability that the next transition is in state $j$ given that at current time the process entered in state $i$ and the index process is equal to $v$. It is simple to realize that
\begin{equation}
p_{ij}^{\lambda}(v)=\lim_{t\rightarrow \infty}Q_{ij}^{\lambda}(v;t).
\end{equation}
\indent Let $H_{i}^{\lambda}(v;\cdot)$ be the sojourn time cumulative distribution in state $i\in E$:
\begin{equation}
\label{hdef}
H_{i}^{\lambda}(v;t):= \mathbb{P}[ T_{n+1}-T_{n} \leq t |  J_n=i,\, U_{n}^{\lambda}=v ]= \sum_{j\in E}Q_{ij}^{\lambda}(v;t).
\end{equation}
\indent It expresses the probability to make a transition from state $i$ with sojourn time less or equal to $t$ given the indexed process is $v$.\\  \indent The conditional waiting time distribution function $G$ expresses the following probability:
\begin{equation}
\label{G}
G_{ij}^{\lambda}(v;t):=\mathbb{P}[T_{n+1}-T_{n}\leq t \mid J_{n}=i, J_{n+1}=j,U_{n}^{\lambda}=v].
\end{equation}
\indent It is simple to establish that
\begin{eqnarray}
&&G_{ij}^{\lambda}(v;t)=\left\{
                \begin{array}{cl}
                       \ \frac{Q_{ij}^{\lambda}(v;t)}{p_{ij}^{\lambda}(v)}  &\mbox{if $p_{ij}^{\lambda}(v)\neq 0$}\\
                         1  &\mbox{if $p_{ij}^{\lambda}(v)=0$}.\\
                   \end{array}
             \right.
\end{eqnarray}
\indent In the papers D'Amico (2011) and D'Amico and Petroni (2012b) explicit renewal-type equations were given to describe the probabilistic behavior of the ISMC model. Similar results could be derived for the WISMC model but here we prefer to derive others results which are of strict relevance to the multivariate model presented in next section.\\
\indent As it is well known, it is possible to give an alternative description of the semi-Markov process by introducing the backward recurrence time process $B(t):=t-T_{N(t)}$ and to describe the probabilistic behavior of the Markov process $(Z(t),B(t))$ on the extended state space $E\times \overline{\N}$ where $\overline{\N}=\{0,1,...,N\}$ and $N$ is the maximum length of stay of the states of the process. This technique was first proposed in Vassiliou and Papadopoulou (1992) and proved useful is studying certain aspects of non-homogeneous semi-Markov process. Also in our more general setting it is possible to describe the system behavior by using the backward recurrence time process, this choice is adopted here to have a description of the one-step transition probabilities of the WISMC model and result to be very useful in the next section for the definition of the bivariate model.\\
\indent Let denote by 
\begin{equation}
\label{pzb}
p_{((i,u)(j,d))}(v):=\mathbb{P}[Z(n+1)=j,  B(n+1)=d \mid Z(n)=i , B(n)=u, U^{\lambda}(n)=v].
\end{equation}
\indent The probabilities $(\ref{pzb})$ can be obtained from the indexed semi-Markov kernel, to proove this, we first need to give the following 
\lemma
\label{lemmaG}
Let suppose that $U^{\lambda}(n)=v$, $T_{N(n)}=n-u$ and $T_{N(n)+1}>n$, then 
\begin{equation}
\label{lemma1}
U_{N(n)}^{\lambda}=v-\sum_{a=T_{N(n)}}^{n-1}f(J_{N(n)},n-a,\lambda)+\sum_{k=0}^{N(n)-1}\sum_{a=T_{N(n)-1-k}}^{T_{N(n)-k}-1}\Delta f(J_{N(n)-k},T_{N(n)},n,a)
\end{equation}
\noindent where $\Delta f(i,T_{N(n)},n,a):=f(i,T_{N(n)}-a,\lambda)-f(i,n-a,\lambda)$. 
\begin{proof}
Let consider the quantity $U_{N(n)}^{\lambda}-U^{\lambda}(n)$. Since $T_{N(n)}=n-u$ and $T_{N(n)+1}>n$, the time $n$ is a waiting time and consequently 
$U^{\lambda}(n)=\sum_{k=0}^{N(n)}\sum_{a=T_{N(n)-k}}^{(n \wedge T_{N(n)-k+1})-1}f(J_{N(n)-k},n-a,\lambda)$. Then
\begin{equation*}
\begin{aligned}
&U_{N(n)}^{\lambda}-U^{\lambda}(n) =\sum_{k=0}^{N(n)-1}\sum_{a=T_{N(n)-1-k}}^{T_{N(n)-k}-1} f(J_{N(n)-1-k},T_{N(n)}-a,\lambda)- \sum_{k=0}^{N(n)}\sum_{a=T_{N(n)-k}}^{(n \wedge T_{N(n)-k+1})-1}f(J_{N(n)-k},n-a,\lambda)\\
&= \sum_{k=0}^{N(n)-1}\!\!\Big( \sum_{a=T_{N(n)-1-k}}^{T_{N(n)-k}-1} \!\!f(J_{N(n)-1-k},T_{N(n)}-a,\lambda)- \!\!\!\!\!\!\!\!\sum_{a=T_{N(n)-k}}^{(n \wedge T_{N(n)-k+1})-1}\!\!\!\!\!\!\!\!f(J_{N(n)-k},n-a,\lambda) \Big)-\sum_{a=T_{0}}^{T_{1}-1}f(J_{0},n-a,\lambda),
\end{aligned}
\end{equation*}
and by considering that $\Delta f(i,T_{N(n)},n,a):=f(i,T_{N(n)}-a,\lambda)-f(i,n-a,\lambda)$ and $U^{\lambda}(n)=v$ by substitution we recover formula $(\ref{lemma1})$
\end{proof}

\theorem
\label{onestep}
For all $i,j \in E$, $u,d\in \N$ and $v\in R$, the one step transition probabilities $p_{((i,u)(j,d))}(v):=\mathbb{P}[Z(n+1)=j,  B(n+1)=d \mid Z(n)=i , B(n)=u, U^{\lambda}(n)=v]$ are given by
\begin{eqnarray}
\label{formula}
&&p_{((i,u)(j,d))}(v)=\left\{
                \begin{array}{cl}
                       \ \frac{\bar{H}_{i}^{\lambda}(v+\Delta U(N(n),n);1+u)}{\bar{H}_{i}^{\lambda}(v+\Delta U(N(n),n);1+u)}  &\mbox{if $j=i,\,\,d=1+u$}\\
                         \frac{{q}_{ij}^{\lambda}(v+\Delta U(N(n),n);1+u)}{\bar{H}_{i}^{\lambda}(v+\Delta U(N(n),n);1+u)}  &\mbox{if $j\neq i,\,\,d=0$}.\\
                   \end{array}
             \right.
\end{eqnarray}
where $\bar{H}_{i}^{\lambda}(t)=1-{H}_{i}^{\lambda}(t)$ is the survival function of sojourn time in state $i$, $q_{ij}^{\lambda}(x,t)=Q_{ij}^{\lambda}(x,t)-Q_{ij}^{\lambda}(x,t-1)$ and $\Delta U(N(n),n)=U_{N(n)}^{\lambda}-U_{n}^{\lambda}$ is the opposit of the variation of the index process on the waiting time $n-N(n)$.
\begin{proof}
Being the events $\{T_{N(n)+1}=k\}$ disjoint it follows that:
\begin{equation}
\label{prob}
\begin{aligned}
& \mathbb{P}[Z(n+1)=j,  B(n+1)=d \mid Z(n)=i , B(n)=u, U^{\lambda}(n)=v]\\
& \mathbb{P}[J_{N(n+1)}=j,  T_{N(n+1)}=n+1-d ,  T_{N(n)+1}>n+1 \mid J_{N(n)}=i, T_{N(n)}=n-u , T_{N(n)+1}>n, U^{\lambda}(n)=v]\\
& +\mathbb{P}[J_{N(n+1)}=j,  T_{N(n+1)}=n+1-d ,  T_{N(n)+1}\leq n+1 \mid J_{N(n)}=i, T_{N(n)}=n-u , T_{N(n)+1}>n, U^{\lambda}(n)=v]
\end{aligned}
\end{equation}
If we represent $U_{N(n)}^{\lambda}=U^{\lambda}(n)+\Delta U(N(n),n)$, the first addend on the r.h.s. of $(\ref{prob})$ becomes
\begin{equation}
\begin{aligned}
\label{prob1}
& \mathbb{P}[J_{N(n+1)}=j,  T_{N(n+1)}=n+1-d ,  T_{N(n)+1}>n+1 \mid J_{N(n)}=i, T_{N(n)}=n-u , T_{N(n)+1}>n, U^{\lambda}(n)=v]\\
& =\mathbb{P}[J_{N(n+1)}=j,  T_{N(n+1)}=n+1-d ,  T_{N(n)+1}> n+1 \mid J_{N(n)}=i, T_{N(n)}=n-u , T_{N(n)+1}>n, U_{N(n)}^{\lambda}=v+\Delta U(N(n),n)]\\
& = \frac{\mathbb{P}[J_{N(n+1)}=j,  T_{N(n+1)}=n+1-d ,  T_{N(n)+1}> n+1, T_{N(n)+1}>n  \mid J_{N(n)}=i, T_{N(n)}=n-u , U_{N(n)}^{\lambda}=v+\Delta U(N(n),n)]}{\mathbb{P}[ T_{N(n)+1}>n  \mid J_{N(n)}=i, T_{N(n)}=n-u ,  U_{N(n)}^{\lambda}=v+\Delta U(N(n),n)]}.
\end{aligned}
\end{equation}
 The denominator of $(\ref{prob1})$ can be computed as follows:
\begin{equation}
\begin{aligned}
\label{denominator}
& = \mathbb{P}[ T_{N(n)+1}- T_{N(n)} >n-(n-u)  \mid J_{N(n)}=i, T_{N(n)}=n-u ,  U_{N(n)}^{\lambda}=v+\Delta U(N(n),n)]\\
& = \mathbb{P}[ T_{N(n)+1}- T_{N(n)} >u  \mid J_{N(n)}=i, T_{N(n)}=n-u ,  U_{N(n)}^{\lambda}=v+\Delta U(N(n),n)]=1-H_{i}^{\lambda}(v+\Delta U(N(n),n); u)
\end{aligned}
\end{equation}
where the last equality is obtained using $(\ref{hdef})$.\\
\indent The numerator of $(\ref{prob1})$ can be evaluated as follows:
\begin{equation}
\begin{aligned}
\label{numerator}
& =\mathbb{P}[J_{N(n+1)}=j,  T_{N(n+1)}=n+1-d ,  T_{N(n)+1}> n+1 \mid J_{N(n)}=i, T_{N(n)}=n-u , U_{N(n)}^{\lambda}=v+\Delta U(N(n),n)]\\
& =\mathbb{P}[J_{N(n+1)}=j,  T_{N(n+1)}=n+1-d \mid T_{N(n)+1}> n+1, J_{N(n)}=i, T_{N(n)}=n-u , U_{N(n)}^{\lambda}=v+\Delta U(N(n),n)]\\
& \cdot \mathbb{P}[ T_{N(n)+1}> n+1\mid J_{N(n)}=i, T_{N(n)}=n-u , U_{N(n)}^{\lambda}=v+\Delta U(N(n),n)]\\
\end{aligned}
\end{equation}
\indent Now note that if $T_{N(n)+1}> n+1$ then $N(n+1)=N(n)$ which implies that $T_{N(n+1)}=T_{N(n)}$ i.e. $m+1-d=n-u$ which gives $d=1+u$. The equality  $N(n+1)=N(n)$ implies also $J_{N(n+1)}=J_{N(n)}$ i.e. $j=i$.\\
\indent Then $(\ref{numerator})$ is equal to
\begin{equation}
\begin{aligned}
\label{numerator1}
& =1_{\{j=i\}}1_{\{d=1+u\}} \mathbb{P}[ T_{N(n)+1}> n+1\mid J_{N(n)}=i, T_{N(n)}=n-u , U_{N(n)}^{\lambda}=v+\Delta U(N(n),n)]\\
& =1_{\{j=i\}}1_{\{d=1+u\}} \mathbb{P}[ T_{N(n)+1}-T_{N(n)}> n+1-(n-u) \mid J_{N(n)}=i, T_{N(n)}=n-u , U_{N(n)}^{\lambda}=v+\Delta U(N(n),n)]\\
& =1_{\{j=i\}}1_{\{d=1+u\}} \big(1-H_{i}^{\lambda}(v+\Delta U(N(n),n); 1+u)\big).
\end{aligned}
\end{equation}
\indent Summarizing $(\ref{prob1})$ is given by
\begin{equation}
\label{addendo1}
\frac{1_{\{j=i\}}1_{\{d=1+u\}} \big(1-H_{i}^{\lambda}(v+\Delta U(N(n),n); 1+u)\big)}{1-H_{i}^{\lambda}(v+\Delta U(N(n),n); u)}.
\end{equation}
\indent It remains to compute the second addend on the r.h.s. of equation $(\ref{prob})$. This probability can be factorized into
\begin{equation}
\label{prob2}
\begin{aligned}
& \mathbb{P}[J_{N(n+1)}=j,  T_{N(n+1)}=n+1-d \mid  J_{N(n)}=i, T_{N(n)+1}\leq n+1,  T_{N(n)}=n-u , T_{N(n)+1}>n, U^{\lambda}(n)=v]\\
& \cdot \mathbb{P}[T_{N(n)+1}\leq n+1 \mid J_{N(n)}=i, T_{N(n)}=n-u , T_{N(n)+1}>n, U^{\lambda}(n)=v]
\end{aligned}
\end{equation}
\begin{equation}
\begin{aligned}
& 1_{\{d=0\}}\Big(\frac{\mathbb{P}[J_{N(n)+1}=j,  T_{N(n)+1}=n+1 \mid  J_{N(n)}=i, T_{N(n)}=n-u , U_{N(n)}^{\lambda}=v+\Delta U(N(n),n)]}{\mathbb{P}[T_{N(n)+1}=n+1 \mid  J_{N(n)}=i, T_{N(n)}=n-u , U_{N(n)}^{\lambda}=v+\Delta U(N(n),n)]}\Big)\\
& \cdot \mathbb{P}[T_{N(n)+1}\leq n+1 \mid J_{N(n)}=i, T_{N(n)}=n-u , T_{N(n)+1}>n, U^{\lambda}(n)=v]
\end{aligned}
\end{equation}
and since $d=0$, we should have $j\neq i$. Then we get
\begin{equation}
\begin{aligned}
& 1_{\{d=0\}}1_{\{j\neq i\}}\Big(\frac{q_{ij}^{\lambda}(v+\Delta U(N(n),n);1+u)}{\mathbb{P}[T_{N(n)+1}=n+1 \mid  J_{N(n)}=i, T_{N(n)}=n-u , U_{N(n)}^{\lambda}=v+\Delta U(N(n),n)]}\Big)\\
& \cdot \Big(\frac{\mathbb{P}[T_{N(n)+1}=n+1 \mid  J_{N(n)}=i, T_{N(n)}=n-u , U_{N(n)}^{\lambda}=v+\Delta U(N(n),n)]}{\mathbb{P}[T_{N(n)+1}> n \mid J_{N(n)}=i, T_{N(n)}=n-u , U_{N(n)}^{\lambda}=v+\Delta U(N(n),n)]}\Big)
\end{aligned}
\end{equation}
\begin{equation}
\label{addendo2}
=1_{\{d=0\}}1_{\{j\neq i\}}\frac{q_{ij}^{\lambda}(v+\Delta U(N(n),n);1+u)}{1-H_{i}^{\lambda}(v+\Delta U(N(n),n))}.
\end{equation}
\indent A substitution of $(\ref{addendo1})$ and $(\ref{addendo2})$ in $(\ref{prob})$ completes the proof.
\end{proof}

\remark
The computation of the probabilities $(\ref{pzb})$ can be done through formula $(\ref{formula})$ where it is necessary to evaluate the quantity $\Delta U(N(n),n)$. This last quantity is obtained thanks to Lemma $(\ref{lemmaG})$ and has to be recalculated step by step.

\section{The Bivariate Weighted-Indexed Semi-Markov Model} 

In this section we extend the WISMC model in a multivariate setting. For reasons of simplicity we will explain the model only for the bivariate case, the multivariate extension is straightforward.\\
\indent Let us assume to dispose of a bivariate series of high-frequency financial data concerning stock returns. Moreover we assume that each one of the two stocks is modeled via a WISMC model. By $J_{n}^{i}$, $T_{n}^{i}$, $U_{n}^{\lambda_{i}}$ and $Z^{i}(n)$ we denote the return at the n-$th$ change, the time of the n-$th$ change, the value of the index at the $n$-th transition and the state of the return at time $t$ for the stock $i\in \{1,2\}$, respectively.\\
\indent In order to define a bivariate model, it is convenient to introduce the backward recurrence time process for the stock $i$ defined, for each time $t\in \N$ by $B^{i}(t)=t-T_{N^{i}(t)}$, where $N^{i}(t)$ is the counting process associated to the stock $i$. The reason for the introduction of the backward recurrence time process is that it complements the semi-Markov process to a Markov process on an extended state of space. This simplifies the definition of the bivariate model which can be now conveniently defined in term of the triplet $(Z^{i}(t), B^{i}(t), U^{\lambda_{i}}(t))$.\\
\indent To define the model we need to formulate three assumptions named in the following A1, A2 and A3. Before of stating the assumption we introduce some auxiliary notation. By $\mathbf{Z}(n)=(Z^{1}(n), Z^{2}(n))$, $\mathbf{B}(n)=(B^{1}(n), B^{2}(n))$, $\mathbf{U}^{\lambda}(n)=(U^{\lambda_{1}}(n), U^{\lambda_{2}}(n))$, $\mathbf{j}=(j_{1},j_{2})$, $\mathbf{i}=(i_{1},i_{2})$, $\mathbf{d}=(d_{1},d_{2})$ and $\mathbf{u}=(u_{1},u_{2})$.\\

ASSUMPTION  A1:
\begin{equation}
\label{A1}
\begin{aligned}
& \mathbb{P}[\mathbf{Z}(n+1)=\mathbf{j},  \mathbf{B}(n+1)=\mathbf{d} \mid \sigma(\mathbf{Z}(h), \mathbf{B}(h)) , 0\leq h\leq n, \mathbf{Z}(n)=\mathbf{i} , \mathbf{B}(n)=\mathbf{u}]\\
& \mathbb{P}[\mathbf{Z}(n+1)=\mathbf{j},  \mathbf{B}(n+1)=\mathbf{d} \mid \mathbf{Z}(n)=\mathbf{i} , \mathbf{B}(n)=\mathbf{u}, \mathbf{U}^{\lambda}(n)=\mathbf{v}]
\end{aligned}
\end{equation}
Assumption A1 states that the knowledge of $(\mathbf{Z}(n)=\mathbf{i}, \mathbf{B}(n)=\mathbf{u}, \mathbf{U}^{\lambda}(n)=\mathbf{v})$ suffices to give the conditional distribution of the couple $(\mathbf{Z}(n+1), \mathbf{B}(n+1))$ whatever the values of the past variables might be.\\
\indent It is simple to realize that:
\begin{equation}
\label{A1bis}
\begin{aligned}  
& \mathbb{P}[\mathbf{Z}(n+1)=\mathbf{j},  \mathbf{B}(n+1)=\mathbf{d} \mid \mathbf{Z}(n)=\mathbf{i} , \mathbf{B}(n)=\mathbf{u}, \mathbf{U}^{\lambda}(n)=\mathbf{v}]\\
& = \mathbb{P}[Z^{1}(n+1)=j_{1},  B^{1}(n+1)=d_{1} \mid Z^{2}(n+1)=j_{2},  B^{2}(n+1)=d_{2} , \mathbf{Z}(n)=\mathbf{i} , \mathbf{B}(n)=\mathbf{u}, \mathbf{U}^{\lambda}(n)=\mathbf{v}]\\
& \cdot \mathbb{P}[Z^{2}(n+1)=j_{2},  B^{2}(n+1)=d_{2} \mid \mathbf{Z}(n)=\mathbf{i} , \mathbf{B}(n)=\mathbf{u}, \mathbf{U}^{\lambda}(n)=\mathbf{v}].
\end{aligned}
\end{equation}
\indent To compute $(\ref{A1bis})$ we need to formulate additional hypotheses:\\

ASSUMPTION A2:
\begin{equation}
\label{A2}
\begin{aligned}  
& \mathbb{P}[Z^{2}(n+1)=j_{2},  B^{2}(n+1)=d_{2} \mid \mathbf{Z}(n)=\mathbf{i} , \mathbf{B}(n)=\mathbf{u}, \mathbf{U}^{\lambda}(n)=\mathbf{v}].\\
& = \mathbb{P}[Z^{2}(n+1)=j_{2},  B^{2}(n+1)=d_{2} \mid Z^{2}(n)=i_{2} , B^{2}(n)=u_{2}, U^{\lambda_{2}}(n)=v_{2}].=:p_{(i_{2},u_{2})((j_{2},d_{2}))}^{2}(v_{2})
\end{aligned}
\end{equation}
\indent The assumption A2 affirms that next state of return and next duration of the stock $2$ do depend only on the same variables at the previous time. This hypothesis can be also considered as a hierarchical assumption: the stock 2 is the leading stock evolving with its own dynamics whereas the stock 1's evolution depends on that of stock 2. It should be noted that it is possible to invert the hierarchy between the two stocks.\\

ASSUMPTION A3:
\begin{equation}
\label{A3}
\begin{aligned}  
& \mathbb{P}[Z^{1}(n+1)=j_{1},  B^{1}(n+1)=d_{1} \mid Z^{2}(n+1)=j_{2},  B^{2}(n+1)=d_{2} , \mathbf{Z}(n)=\mathbf{i} , \mathbf{B}(n)=\mathbf{u}, \mathbf{U}^{\lambda}(n)=\mathbf{v}]\\
& = \mathbb{P}[Z^{1}(n+1)=j_{1},  B^{1}(n+1)=d_{1} \mid sgn(Z^{2}(n+1))=s, Z^{1}(n)=i_{1},  B^{1}(n)=u_{1} , U^{\lambda_{1}}(n)=v_{1}]\\
& =:\tilde{p}_{(i_{1},u_{1})((j_{1},d_{1}))}^{1}(v_{1};s)
\end{aligned}
\end{equation}
\noindent where $sgn(Z^{2}(n+1)))$ is the sign of $Z^{2}(n+1)$ which can assume the values $+, 0, -$ according to the fact that the stock 2 exhibits a positive, constant or negative return, respectively. This assumption is very important as it reduces drastically the dimensionality of the model still preserving the cross correlation between the two stocks.\\
\indent Summarizing, the assumptions A1, A2 and A3 allow us to compute the joint one step transition probability $(\ref{A1})$ of the two stocks with the product 
\begin{equation}
\tilde{p}_{(i_{1},u_{1})((j_{1},d_{1}))}^{1}(v_{1};s)p_{(i_{2},u_{2})((j_{2},d_{2}))}^{2}(v_{2}).
\end{equation}
\indent The probabilities $p_{(i_{2},u_{2})((j_{2},d_{2}))}^{2}(v_{2})$ have been evaluated in the previous section where they were represented as a function of the weighted-indexed semi-Markov kernel. In the paper D'Amico, Petroni and Prattico (2013) a nonparametric estimator of the weighted-indexed semi-Markov kernel was derived. From this estimator it is immediate to recover a plug-in estimator of $(\ref{prob})$.\\
\indent The probabilities $\tilde{p}_{(i_{1},u_{1})((j_{1},d_{1}))}^{1}(v_{1};s)$ can be also evaluated directly from the data. To this end it is sufficient to consider the estimator $\frac{N_{L}^{1,2}(i_{1},u_{1},v_{1};j_{1},d_{1},s_{2})}{N_{L}^{1,2}(i_{1},u_{1},v_{1};s_{2})}$ where $L$ is the lenght of the bivariate series of stock returns and   
$$N_{L}^{1,2}(i_{1},u_{1},v_{1};j_{1},d_{1},s_{2})=\sum_{t=1}^{L}1_{\{  Z^{1}(t)=j_{1},  B^{1}(t)=d_{1} ,sgn(Z^{2}(t))=s_{2}, Z^{1}(t-1)=i_{1},  B^{1}(t-1)=u_{1} , U^{\lambda_{1}}(t-1)=v_{1} \}}$$  
and
$$N_{L}^{1,2}(i_{1},u_{1},v_{1};s_{2})=\sum_{j_{1}\in E}\sum_{d_{1}\in \bar{\N}}N^{1,2}(i_{1},u_{1},v_{1};j_{1},d_{1},s_{2}).$$

\section{Empirical results}
The model described in the previous sections was applied to a set of 20 stocks from the Italian Stock Exchange (``Borsa Italiana"). The list of stocks and their symbols are reported in table \ref{tab}. 
\begin{table}
\begin{center}
\begin{tabular}{|l|l|}
\hline
\textbf{AT}&Atlantia\\\hline
\textbf{MP}&Banca Monte dei Paschi di Siena\\\hline
\textbf{E}&ENI\\\hline
\textbf{EN}&ENEL\\\hline
\textbf{F}&Fiat\\\hline
\textbf{FN}&Finmeccanica\\\hline
\textbf{G}&Generali\\\hline
\textbf{IS}&Intesa San Paolo\\\hline
\textbf{LU}&Luxottica\\\hline
\textbf{MS}&Mediaset\\\hline
\textbf{MB}&Mediobanca\\\hline
\textbf{PC}&Pirelli\\\hline
\textbf{PR}&Prysmian\\\hline
\textbf{SP}&Saipem\\\hline
\textbf{SR}&Snam Rete Gas\\\hline
\textbf{ST}&ST Microelectronics\\\hline
\textbf{TI}&Telecom\\\hline
\textbf{TE}&Tenaris\\\hline
\textbf{TR}&Terna\\\hline
\textbf{UC}&Unicredit\\\hline
\end{tabular}
\end{center}
\caption{Stocks used in the application and their symbols}\label{tab}
\end{table}
The database is composed of tick-by-tick quotes recorded form January 2007 to December 2010 (4 full years). The data have been re-sampled to have 1 minute frequency. The number of returns analyzed is then roughly $500*10^3$ for each stock. A better description of the database can be found in D'Amico \& Petroni (2011).
Returns have been discretized into 5 states chosen to be symmetrical with respect to returns equal zero and to keep the shape of the distribution unchanged. Returns are in fact already discretized in real data due to the discretization of stock prices which is fixed by each stock exchange and depends on the value of the stock. Just to make an example, in the Italian stock market for stocks with value between 5.0001 and 10 euros the minimum variation is fixed to 0.005 euros (usually called tick). We then tried to remain as close as possible to this discretization. 

\begin{figure}
\centering
\includegraphics[height=8cm]{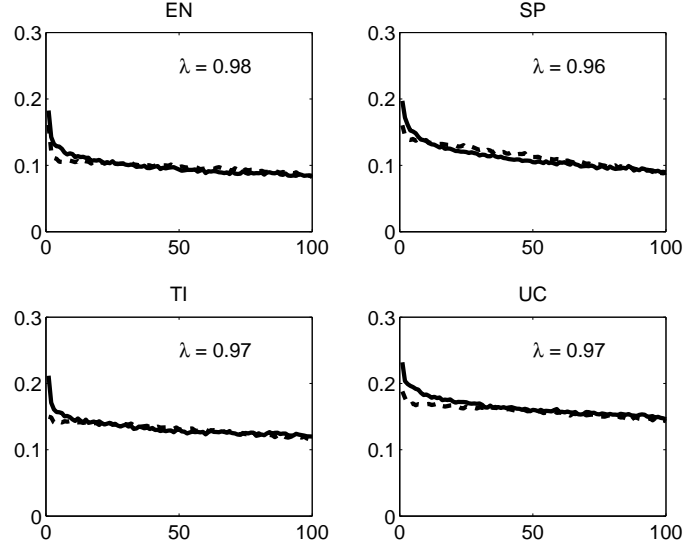}
\caption{Autocorrelation functions of real data (solid line) and synthetic (dashed line) time series for the analyzed stocks.}\label{fig1}
\end{figure}

Following D'Amico \& Petroni (2012b) we use as definition of the function $f$ in  (\ref{funcrela}) an exponentially weighted moving average (EWMA) of the squares of returns which has the following expression:
\begin{equation}
\label{funct}
f(J_{n-1-k},a,\lambda)=\frac{\lambda^{T_{n}-a} J_{n-1-k}^2}{\sum_{k=0}^{n-1}\sum_{a=T_{n-1-k}}^{T_{n-k}-1}\lambda^{T_{n}-a}}
\end{equation}
\noindent and consequently the index process becomes
\begin{equation}
\label{ewma}
U_{n}^{\lambda}=\sum_{k=0}^{n-1}\sum_{a=T_{n-1-k}}^{T_{n-k}-1}\Bigg(\frac{\lambda^{T_{n}-a} J_{n-1-k}^2}{\sum_{k=0}^{n-1}\sum_{a=T_{n-1-k}}^{T_{n-k}-1}\lambda^{T_{n}-a}}\Bigg).
\end{equation}
The index $U^\lambda$ was also discretized into 5 states of low, medium low, medium, medium high and high volatility.
Using these definitions and discretizations we estimated, for each stock, the probabilities defined in the previous section by using their estimators directly from real data. 
By means of Monte Carlo simulations we were able to produce, for each of the 20 stocks, a synthetic time series. Each time series is a realization of the stochastic process described in the previous section with the same time length as real data.
Statistical features of these synthetic time series are then compared with the statistical features of real data. In particular, we tested our model for the ability to reproduce the autocorrelation functions and the cross-correlation betweens stocks.  
We remind the definition of the autocorrelation function: if $R$ indicates returns, the time lagged $(\tau)$ autocorrelation of the square of returns is defined as 
\begin{equation}
\label{autosquare}
\Sigma(\tau)=\frac{Cov(R^2(t+\tau),R^2(t))}{Var(R^2(t))}
\end{equation}
We estimated $\Sigma(\tau)$ for real data and for synthetic data and show in Figure \ref{fig1} a comparison between them for 4 stocks chosen from the 20 stocks in the database. 

It is possible to note that our model is able to reproduce almost perfectly the autocorrelation behavior of these stocks. Note that each stock has its own best value for the parameter $\lambda$. These results were already described in D'Amico \& Petroni (2012b) for different stcks. We want to test here also if the bivariate model is able to reproduce the pairwise correlation (cross-correlation) between stocks still preserving the autocorrelation functions. The definition of the cross-correlation between stocks $\alpha$ and $\beta$ is: 
\begin{equation}
\label{autosquare}
\Sigma_{\alpha,\beta}=\frac{Cov(R_{\alpha},R_{\beta})}{\sqrt{Var(R_{\alpha})Var(R_{\beta})}}
\end{equation}
From the real time series and from the synthetic ones we estimated the cross-correlation matrix for each couple of stock. Note that the matrix is symmetric with respect to stocks $\alpha$ and $\beta$. We then report in the table only a lower triangular matrix.

\begin{table}
\begin{center}
\begin{tabular}{||l|*{20}{r}c}
\textbf{MP}&13\\
\textbf{E}&14&14\\
\textbf{EN}&16&16&26\\
\textbf{F}&15&15&22&27\\
\textbf{FN}&14&14&16&19&18\\
\textbf{G}&16&17&25&29&25&190\\
\textbf{IS}&15&17&23&27&25&18&27\\
\textbf{LU}&13&13&15&16&16&14&17&16\\
\textbf{MS}&14&14&17&19&18&15&19&18&15\\
\textbf{MB}&14&16&16&18&17&15&20&19&15&16\\
\textbf{PC}&9&10&10&11&11&10&12&11&10&10&11\\
\textbf{PR}&11&11&13&14&13&12&14&14&11&12&12&9\\
\textbf{SP}&15&15&19&24&21&16&22&21&15&17&17&10&13\\
\textbf{SR}&11&10&14&14&12&12&14&12&11&12&11&7&9&12\\
\textbf{ST}&15&15&19&22&21&17&22&21&16&17&17&11&13&19&13\\
\textbf{TI}&12&13&18&22&19&14&21&19&13&15&14&9&11&16&11&17\\
\textbf{TE}&15&15&19&23&21&17&22&21&16&17&17&11&14&21&12&20&17\\
\textbf{TR}&9&9&13&12&11&10&12&11&10&10&10&7&8&11&12&11&10&11\\
\textbf{UC}&15&17&23&28&27&18&28&30&16&18&19&11&14&21&12&22&20&22&11\\\hline
&\textbf{AT}&\textbf{MP}&\textbf{E}&\textbf{EN}&\textbf{F}&\textbf{FN}&\textbf{G}&\textbf{IS}&\textbf{LU}&\textbf{MS}&\textbf{MB}&\textbf{PC}&\textbf{PR}&\textbf{SP}&\textbf{SR}&\textbf{ST}&\textbf{TI}&\textbf{TE}&\textbf{TR}\\\hline\hline
\end{tabular}
\caption{Cross-correlation matrix (multiplied by 100) for real data.}\label{c1}
\end{center}
\end{table}

\begin{table}
\begin{center}
\begin{tabular}{||l|*{20}{r}c}
\textbf{MP}&6\\
\textbf{E}&9&8\\
\textbf{EN}&9&9&13\\
\textbf{F}&9&9&11&13\\
\textbf{FN}&7&8&8&8&8\\
\textbf{G}&10&11&12&14&13&10\\
\textbf{IS}&7&8&9&10&10&8&11\\
\textbf{LU}&6&7&6&7&7&7&7&7\\
\textbf{MS}&7&8&8&8&8&8&9&9&8\\
\textbf{MB}&7&9&7&8&8&8&9&10&8&8\\
\textbf{PC}&4&4&3&4&4&4&4&4&4&4&5\\
\textbf{PR}&5&6&5&5&5&6&6&6&6&5&6&6\\
\textbf{SP}&8&8&9&11&10&9&10&11&9&9&9&7&8\\
\textbf{SR}&5&5&6&6&5&6&6&6&5&6&5&4&5&5\\
\textbf{ST}&8&9&9&10&10&8&10&11&9&9&9&8&8&9&7\\
\textbf{TI}&7&7&8&9&9&7&9&9&7&7&7&6&7&7&6&8\\
\textbf{TE}&8&8&9&10&10&8&10&11&9&9&9&8&9&10&7&10&9\\
\textbf{TR}&4&4&4&4&4&4&4&4&4&4&4&4&3&4&5&4&4&4\\
\textbf{UC}&8&10&10&12&13&9&12&15&9&9&10&9&9&10&6&11&10&11&7\\\hline
&\textbf{AT}&\textbf{MP}&\textbf{E}&\textbf{EN}&\textbf{F}&\textbf{FN}&\textbf{G}&\textbf{IS}&\textbf{LU}&\textbf{MS}&\textbf{MB}&\textbf{PC}&\textbf{PR}&\textbf{SP}&\textbf{SR}&\textbf{ST}&\textbf{TI}&\textbf{TE}&\textbf{TR}\\\hline\hline
\end{tabular}
\caption{Cross-correlation matrix (multiplied by 100) for synthetic data.}\label{c2}
\end{center}
\end{table}
From the two tables reported here it is possible to note that our bivariate model is able to reproduce more than $50\%$ of the cross-correlation. In our view this is a good results given that the dependence between stocks is modeled in a very simple way.

\section{Concluding remarks} 

With the aim to reproduce cross-correlation between stocks and following our previous works on univariate returns model, we have modeled financial price changes through a bivariate weighted indexed semi-Markov model. 

The results presented here show that the semi-Markov kernel is influenced by the past volatility and that its influence decrease exponentially with time. In fact, if the past volatility is used as an exponentially weighted index, the model is able to reproduce almost exactly the behavior of market returns: the returns generated by the model are uncorrelated while the square of returns present a long range correlation very similar to that of real data. Moreover the generalization to bivariate process, even if very simple, is able to reproduce more than $50\%$ of the real cross-correlation.

We stress that our model is very different from those of the ARCH/GARCH family. We do not model directly the volatility as a correlated process. We model returns and by considering the semi-Markov kernel conditioned by a weighted index the volatility correlation comes out freely.

\end{document}